\documentclass{article}
\usepackage{graphicx}
\begin{document}
\title{Quantum Indeterminism and First Passage Random Walks in Hilbert Space}
\author{Fariel Shafee\footnote{fshafee@alum.mit.edu}\\ Department of Physics\\ Princeton University\\
Princeton, NJ 08540\\ USA.}
\date{}
\maketitle
\begin{abstract}
We propose a new model for a measurement of a characteristic of a microscopic quantum state by a large system that selects stochastically the different eigenstates with appropriate quantum weights. Unlike previous works which formulate a modified Schr\"odinger equation or an explicit modified Hamiltonian, or more complicated mechanisms for reduction and decoherence to introduce transition to classical stochasticity, we propose the novel use of couplings to the environment, and random walks in the product Hilbert space of the combined system, with first passage stopping rules, which seem intuitively simple, as quantum weights and related stochasticity is a commonality that must be preserved under the
widest range of applications, independent of the measured quantity and the specific properties of the measuring device.

\vspace*{0.5cm}
\end{abstract}
]
\section{INTRODUCTION}
Quantum measurement remains one of the most puzzling physical processes.  Although the laws of the quantum world and those of the classical world remain experimentally verifiable, the process of gaining information about the quantum world in the macroscopically perceived world remains unsolved.

Quantum mechanically an object may have various states associated with different values of a property superposed together, and interference patterns support the existence of such superposed quantum waves \cite{AC1}. On the other hand in the classical world, superposed states do not exist, and interference patterns are not observed with of classical matter constituting of many quantum particles.

In an ensemble, the probability of finding an object with a certain well-defined quantum property is equal to the square of the amplitude of the respective component of the  wave function. Hence, in an ensemble of quantum particles such that each has two possible spin states, up and down, with wave amplitudes {\bf a} and {\bf b} superposed together in a quantum world, after a measurement is made in the classical world each particle comes either in the state up or in the state down.  The probability of finding an up particle in such an ensemble is ${\bf |a|^2}$ and the probability of finding a down particle is ${\bf |b|^2}$.  So, in each measurement process only one of the possible quantum states is retrieved and the rest of the superposed states are lost. In the measurement process, therefore, interference patterns are lost with the loss of coherence among quantum states, and only one of many states is obtained in the classical world, so that information about that quantum system is also partially lost.

In recent years there has been renewed interest (review in
\cite{SC1}) in the indeterminism in the measurement process in quantum mechanics, which
has remained an enigmatic probabilistic characteristic of quantum theory since
Bohr and the Copenhagen interpretation \cite{AN1,BU1}. Some people have even seen in the confusion a fuzzy mixing of ontology and epistemology, and the role of consciousness, which continues to be debated \cite{HP1}. Intriguing many-world interpretations with bifurcating realities have also been suggested \cite{MT1}.

In \cite{ZUREK}, it was proposed that it is the many degrees of freedom of the environment that causes the stochastic collapse of the wave function.  Each quantum state finds an environmental state partner, which it prefers, and these pairs get separated, and hence decohered, because of the orthogonality of the environmental states due to many degrees of freedom of the environment.  However, the sudden loss of information and coherence remains confusing.

The existence of these measurer-microsystem entangled systems,  and  the reduction by itself, raises the question of where these lost component states are disappearing and how the environment can have so many orthogonal states corresponding to each quantum state, and how or why the macroscopic states are getting bound to these small quantum states.

In this paper, we propose a new model where stochasticity is derived in a  dynamic fashion and in steps.  Decoherence occurs simultaneously with reduction.  The initial pairing is among quantum and corresponding mesoscopic states.  One of these pairs gets amplified in the detector and locks the detector in an energetically favorable state indicating the existence of that one amplified quantum state.  The rest of the bound pairs are dissipated in the environment or are rotated, so that information about the other quantum states are not expressed in the environment.  The many degrees of freedom associated with the environment causes the dissipated states to be lost for ever in small packets each carrying parts of the information. The couplings of the detector causes the one expressed state to be amplified to a scale where information can be retrieved macroscopically.  So, this new approach introduces dynamics in the amplification and loss of states together with the loss of coherence.

Our model uses random walks to amplify the preferred state within the detector.  Random walks with hierarchical constraints have also been presented recently \cite{OM1} in a different picture to explain reduction to a pure eigenstate. In classical systems there is no intrinsic indeterminism, and random walks are simply an algorithmic approximation to simulate extremely complicated dynamics resulting from a system with a large number of coupled degrees of freedom. In the quantum context too one might expect a similar evolution replicable in the large by a probabilistic interpretation, when a mesoscopic detector interacts with a microscopic system, despite a deterministic set of rules of dynamics.  However, there is as yet no universally acceptable theory, and all existent models have their advantages and weaknesses needing further development.
Our objective is to produce a picture that is more comprehensible and plausible than the others in certain respects.

In the following subsection we shall discuss briefly the mechanism of reduction proposed recently by Omn\'{e}s, and thereafter another interesting method proposed by Sewell \cite{SEW}, before we present details of our work, where we give a general method of picturing and explaining the transition from mixed quantum states to pure eigenstates as first passage walks in Hilbert space. This model also includes a preparation stage, where the system and the detector form a virtual superposed quantum bound state
before the walk begins.

\section{COMPARISON WITH RECENT MODELS}
\subsection{Current Models of Decoherence}
Theories of quantum collapse are usually based on decoherence  or reduction \cite{ZUREK}. In decoherence, a quantum state interacts with the environment, which contains a collection of all possible observable states, and due to this interaction the cross terms in the density matrix ( $|i\rangle \langle j|$, $j\neq i$)  gradually vanish, leaving the arena to the diagonal terms only ($i=j$), but that mixture too finally reduces to one single eigenstate, which corresponds to the result of the measurement.

The quantum system is initially in the state

$|\Psi\rangle = \sum_i |i\rangle \langle i|\Psi>$

when expressed in terms of the $|i\rangle$'s which form the basis selected or preferred by the environment.
If the system is brought into contact with the environment, $|E\rangle$, a joint state results in the form
$\sum_i |i\rangle|E\rangle \langle i|\Psi\rangle$
This compounded system is allowed to interact, so that either the system is lost to the environment,

\begin{equation}\label{eq1}
|i\rangle|E\rangle \rightarrow |E_i\rangle
\end{equation}
 or, the environment is modified, so that the joint system evolves to
 \begin{equation}\label{eq2}
 \sum_i |i,E_i \rangle \langle i|\Psi\rangle
 \end{equation}
 In both cases, the selection of one of many possible quantum states is achieved by assuming the {\bf ``orthogonality"}  of the environment states posed by the many degrees of freedom in the environment
 $<E_i|E_j>~\delta_{ij}$

However, the main criticisms of decoherence remain:
\\
{\bf 1.} The ``einstates'' (environment-selected eigenstates) are inserted in an {\em ad hoc} manner, with no explanation.
\\
{\bf 2.} The splitting of the macro-system into a relevant system and the environment, by means of a set of projection operators, is also done in an {\it ad hoc} manner, without a credible strong procedural explanation. Nevertheless, it is this drastic classification that leads to the loss of coherence.
\\
{\bf 3.} The diagonalization of the reduced density matrix of the relevant system does not offer any insight into the dynamics of the mechanism, and also does not completely eliminate small probabilities.  The problems with diagonalization was pointed out in \cite{OM1}.
\\
{\bf 4.} The orthogonality of the environment states is mentioned. However, what these orthogonal environment states could be physically, and how they exist independent of the quantum system, or what happens to the remaining environment states after one is selected, is not clear.  The fate of the unselected quantum states is also unclear in regard to where they disappear.
\\
\subsection{The Omn\'{e}s Paradigm of Reduction}

Some of the ambiguities posed in the process proposed by \cite{ZUREK} was resolved by Omn\'{e}s.
In order to account for the dynamics of decoherence, in a recent work, Omn\'{e}s \cite{OM1} combined reduction with decoherence.  His model suggested the adjustment of the coefficients of the projection operators for different eigenstates so that an infinitesimal reduction in the j'th state changes the projection operator $P_j$ to $(1 + \epsilon_j)$ $P_j$, in turn changing the probability amplitude of the j'th state.
The random adjustment of the weights was carried out by using homogenous, isotropic Brownian motion.

In his paper, Omn\'{e}s proposed combining two possible explanations of quantum measurement, namely decoherence and also reduction. The addition of reduction adds a timescale, so that the sudden diagonalization problem is solved, and the problems with small probabilities existing in a decohered system is also solved when the reduction of states in each vertex is introduced.

However, although this method shed some light into possible dynamics leading to decoherence, the main criticisms of decoherence, regarding the {\em  ad hoc} nature of the eigenbasis and projection vectors in the environment, and the physical basis of projection vectors readjusting themselves remain unanswered.

In the Omn\'{e}s approach reduction occurs as the last step of decoherence. This necessitates the reduction of already existing multiple states into one state.  Where the reduced decohered states vanish remains unanswered.

\subsection{The Sewell Paradigm}
Sewell proposes using many-body Schr\"{o}dinger equations for the large number of degrees of freedom for the composite of the microsystem and the macroscopic measuring system. This finite  closed system with conservative dynamics with no dissipation is claimed to be sufficient to bring about the collapse of the superposed state to an eigenstate in number of steps.

This method avoids the need of the decoherence process to end eventually in `consciousness', as envisaged in the steps proposed by von Neumann \cite{NEU1} and Wigner \cite{WIG1}. It gives a robust one-to-one correspondence between the microstates of the measured system and the macrostates of the instrument, irrespective of the initial quantum state.

Sewell obtains conditions on the measuring devices imposed by the requirement of obtaining quantum measurement as probabilistic observation. However, one property signifies that the micro-macro coupling removes the interference
between the different components of the pure state and thus represents a complete
decoherence effect.

\subsection{Our Approach}
In this paper, we propose a more physically comprehensible approach, which circumvents some of the mathematical abstractions posed in these previous works. Detailed arguments and steps are inserted to reduce the ambiguities in existing models. Our approach also treats the elimination of cross terms simultaneously with reduction, so that multiple decohered  states do not exist entangled independently with possible orthonormal environment states, to be eliminated one by one, with intrinsic problems of normalization.  We also clarify what these environment states may be and how one state emerges macroscopically while the others are eliminated.  Our approach does not necessitate the existence of an ultimate undefinable detector (consciousness) to measure a state.  The approach also does not make it necessary to introduce parallel universes or multiple universes to explain the disappearance of any unmeasured quantum state.

Couplings between the microscopic world and mesostates are obtained by using interactions, and the coupling constants are taken to be proportional to the amplitudes of the superposed waves. The initial couplings, that due to further couplings within the measuring device, also in the presence of stochastic interactions of the detector subsystems with the environment, are reduced by means of first passage random walk  and reduce the cross terms along with the reduction of multiple states.

In the next part of this paper, we present a method of concurrent decoherence and reduction of quantum information within a coupled detector by using first passage random walks and the formation of images.  In the last part, we discuss some possible methods of such image formation and reduction given the necessity to preserve unitarity.  Some related dynamical mechanisms and philosophical questions are addressed, and the relevance of the model proposed by us, given the degree of ambiguity and incompleteness still existing in the process of quantum measurement, is discussed.


\section{FORMATION OF IMAGES}

Though physical laws are all expected to be based on quantum
principles, the full quantum picture is clear and usable only for
small systems, such as the interaction among a small number of
particles. When a large number of particles are involved,
approximations become inevitable. Even in quantum field theory one
often has to resort to approximate effective interactions to
simulate contributions of the sum of large diagrams involving many
propagators and loops, losing in the process some vital components
of the quantum theory such as unitarity, and making perturbative
convergence suspect.

In a measuring device there will in general be a large number of
subsystems which can directly or indirectly couple to the
attribute of the small system which is to be measured and they
will in turn couple to large macroscopic recording devices whose
states indicating the different values of the measured quantity
are macroscopically so different with such huge energy barriers in
the transition paths that it is not possible to tunnel from one
such state to another in a realistic time limit, so that we do not
expect quantum superpositions of the states of the recording
device after measurement is completed. However, during the process
of measurement the microsystem components of the device (D)
coupling to the measured microsystem (S) may be in states of
superposition of eigenstates. We shall now assume that D and S
couple in a way to form a virtual bound pair, the D state being an
image of the S state in the following sense:

\begin{equation} \label{eq3}
|\psi\rangle_S  = \sum_i a_i |i\rangle_S
\end{equation}

has the image

\begin{equation} \label{eq4}
|\psi\rangle_D  = \sum_i a_i^* |i^*\rangle_D
\end{equation}

We can here refer to a comparison with the creation of a
conjugate image charge in a grounded neutral conductor, which has a
sea of charges of both kinds available. When a free charge
approaches it, there is an induction of the image charge, which may
be a manifestation of a re-arranged charge distribution on the
conductor, and not of any particular real charge on the conductor.
So $|\psi\rangle_D$ may be the effective state resulting from
the combination of a large number of micro subsystem components of
the device. Hence, the image states are not exact clones of the
micro-system's states, and they correspond on a class-to-one basis, each image class containing a superposition of a large number of quantum states, not discriminable on a coarse-grained macro or mesoscale,  different effective unitary operations may apply to form images
for different microstates and the ``no-cloning theorem" \cite{NC1} is not
applicable here. Elsewhere \cite{FS1} we have studied and
demonstrated a possible mechanism of the formation of such conjugate images.

The formation of the image states and the virtual bound pairs can be briefly justified as below (the example can be generalized):
1. Consider detection of spin $s_z$ along the preferred direction of the environment of the detector.  Let the mesoscopic corresponding variable in D be $S_z$.  The interaction energy is $\pm k s_z S_z$.So the lowest energies correspond to $s_z=-1/2$, $S_z=+S$, and $s_z=-1/2$, $S_z=-S$, where S may be a little fuzzy when obtained by coarse graining. If the incoming state is polarized along some other direction initially, it will be expressible as a linear superposition  of the preferred eigenbasis as, say,
\begin{equation}
|in\rangle = a |+1/2\rangle + b |-1/2\rangle
\end{equation}

2. The interaction energy between the system and the detector for the two states would be proportional to $a$ and $b$ respectively.

\subsection{Detector Image Coefficients and Quantum Wave Coefficients}
In the image proposed above, the detector forms an image corresponding to the quantum wave and the states of the detector share the same (conjugate) coefficients as those of the quantum states.
However, the detector itself is a large system comprising of many subsystems.  Hence, the image ``wave" in the detector must comprise of an ensemble of coupled subsystems.  The surface area or number of subsystems corresponding to each quantum state (the detector subsystem (meso-system) binding with the corresponding quantum system in an energetically favorable interaction) would reflect the coefficient of the corresponding detector image component.  The cross terms derive from the interactions from energetically unfavorable pairs of mismatched states.  If all the detector subsystems correspond to a single quantum eigenstate, and are coupled together to form a favorable energy minimum, the detector state is locked to indicate that certain quantum state, and since all the states correspond to a single quantum state, the cross-terms also disappear.

We also assume these subsystems in the mesoscopic detector system are coupled to the macroscopic recording part of the device (R) in such a way that when they act in unison they can change the state of R to indicate one of
the eigenvalues of S. The image state of D and the recording state
of R may be degenerate states and superpositions of the corresponding multi-component
meso or macro systems may correspond to unique eigenvalues of S. We do not need to make use of any decoherence arguments in the usual sense.

\section{LOSS OF REDUCED STATES}

In forming the virtual S-D bound system, if conservation rules
demand, excess values of the measured attribute and intrinsically
related quantities may pass on to other subsystems of R, or to the
environment, in the same way as the excess charge from a grounded
conductor passes to the ground after the formation of the image. We can state this as a theorem:

{\bf Theorem}: It is not possible in general for a closed quantum system comprising only the detected system and the detector to perform a complete measurement.

{\em Proof:} Let us consider a microsystem with spin-1/2 given in Eq. 1. This can be generalized to other types of measurement. Let the detector have a preferred basis (i.e. z-direction) different from that used in Eq. 1. After interaction leading to measurement, the micro-system assumes an eigenstate corresponding to the eigenbasis of D, and hence the composite system now has no component of the spin in any direction normal to the eigenbasis direction, even though the original system did have some, in general. Since angular momentum must be conserved, this is not possible. In other words, the component of spin of the microsystem orthogonal to the direction of the eigenbasis of D must escape from the (S-D) quantum system to the environment.

Though in QM we always have the same left-hand side in the  completeness relation of a basis

\begin{equation}
1 = \sum |i\rangle \langle i|
\end{equation}

the basis we choose on the right-hand side must be relevant to the context. In a measurement problem the operative basis set corresponds to the environment in which the detector is placed, and not the original orientation of the quantization axis of the microsystem which is detected. When a two-body problem is reduced to a one-body problem using reduced mass, the origin remains closer to the heavier body. In the quantum measurement problem too, the mesoscale D system consisting of a very large number of microsystems comparable to the detected system S. Though quantum states are normalized to unity, independent of the size of the system, they can also be represented as rays in Hilbert space for many purposes, where absolute probability is not needed. The energies of alignment are quite different and cannot be treated in terms of a size-independent parameter, as in the case of probability. The energy exchanges involved in rotating the alignment axis of S, with respect to the environment (e.g. an external magnetic field) to bring it in line with that of S will be many orders of magnitudes larger than the energy required to for S to align to D. Hence, in a sense, the detector D has a higher quantum ``inertia" and fixes the frame of the basis set for the measurement process.

The S-D coupled system may be represented by

\begin{eqnarray}  \label{eq5}
|\psi\rangle_{SD} = \sum_i |a_i|^2 |i\rangle_S |i^*\rangle_D
+\sum_{i\neq j}a_i a_j^* |i\rangle_S |j^*\rangle_D
\end{eqnarray}

\section{FIRST PASSAGE TRANSITION TO EIGENSTATES}


We now propose that the complicated quantum interactions of the
macroscopic D-R complex and S can be represented by a random walk in
Hilbert space of the (SD) coupled system with active transfers among
the microstate-image pair diagonal terms in Eq. \ref{eq3}, with the
off-diagonal terms, required by unitarity, which are uncoupled free
spectator components adjusting their coefficients passively to
conserve unitarity. When the coupled system finally reaches a
particular eigenstate pair, all such off-diagonal terms vanish. We
do not intend to construct a model of explicit unitary
transformations which would lead to such a scenario. But it is
possible to show easily that a sequence of simple unitary rotations
in Hilbert space through a uniform stepping angle of random sign
cannot achieve the right probabilities even in the case of a qubit,
i.e. a spin-1/2 object. It is probably unrealistic to expect that
the micro-transitions leading eventually to a complete collapse to
an eigenstate is expressible in terms of a calculable sequence of
operations. Indeed each successive operation may be drastically
different from its predecessor. Brownian motion
\cite{PE1,PE2,GH1,DI1} may be the most simplified model for such a
sequence.

We approach the transition as a first passage problem \cite{FP1},
where, on reaching an eigenstate, which forms a bounding wall in
the Hilbert space, the evolution of the system stops, the recorder
shows the eigenvalue, and that eigenstate of S continues until
further measurement. In game theory language it can also be
described as a winner-takes-all game with the players (the
competing eigenstates) betting against one another in pairs by
random turns, with an equal small $1:1$ stake (the interaction) at
each turn, and players eliminated one by one on going bankrupt
till the eventual winner emerges. An incomplete game would
indicate a quantum mixed final state, which may be the case when
the recorder's state is coupled to such a mixture.

For the measurement of a qubit

\begin{equation}  \label{eq6}
|S_{in}\rangle = a_0 |0\rangle + a_1 |1\rangle
\end{equation}

our model yields the compound (SD) state (omitting passive
cross-terms):

\begin{equation}  \label{eq7}
|SD_{in}\rangle = |a_0|^2 |00^*\rangle_{SD} + |a_1|^2
|11^*\rangle_{SD}+\quad \textrm{off-diagonal cross-terms}
\end{equation}

We locate this initial state as the point $(x_0,y_0) = (|a_0|^2,
|a_1|^2)$ in a two-dimensional Hilbert space. We may anticipate
$x+y=1$ here, and eliminate one co-ordinate, but later we shall
keep the co-ordinates free till the last stage of the calculation.

So we have a random diffusion of the probability concentration
from $x_0$ to  $x=0$ (pure eigenstate $|1\rangle$) and to $x=1$
(pure eigenstate $|0\rangle$).

With the diffusion equation

\begin{equation} \label{eq8}
\frac{\partial c }{ \partial t}= D \frac{\partial^2 c}{\partial
x^2}
\end{equation}

where the diffusion constant $D$ is related to an effective strength
of interaction or a length of the walk, and $t$ is a continuous
variable representing the sequence index of  small discrete
operations, which may be proportional to real time. In Fig. 1
we give a diagram of the process. We should, however, not
interpret it as a Feynman-type graph, and should also remember that
the cross-terms also change with every step passively to maintain
unitarity. The situation is to some extent analogous to the
quantization of spin  $s_z$ by an interacting external magnetic
field $H_z$, while  the magnitude of the resultant of the other
components add up vectorially to maintain the constrained value of
the overall magnitude of the spin $s$.

\begin{figure}[ht!]
\begin{center}
\includegraphics[width=10cm]{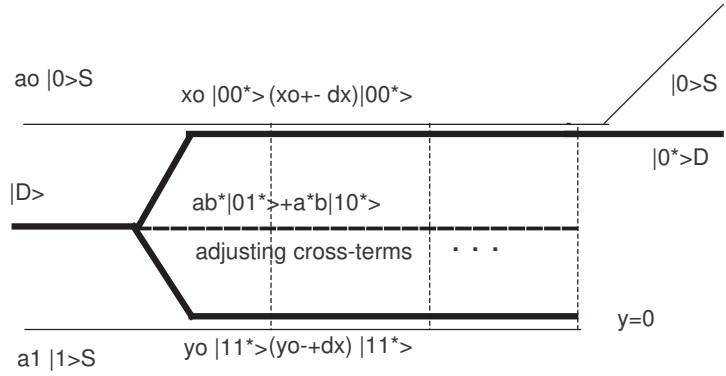}
\end{center}
\caption{\label{fig5025}Successive random transformations on competing
eigenstates  $|00^*\rangle_{SD}$ and $|11^*\rangle_{SD}$ terminating
in the elimination of one.
 }
\end{figure}

Since we shall have to sum over all possible step numbers, and
equivalently integrate over all $t$, it is more convenient to work
with the Laplace transform of $c$, with the transformed equation

\begin{equation}  \label{eq9}
\frac{d^2\tilde{c}(x,s)}{dx^2}-(s/D)\tilde{c}(x,s)= -c(x,t=0)/D
\end{equation}

$s$ being the Laplace conjugate of $t$.

The initial $c(x,0)$ is the delta function $\delta (x-x_0)$ when
the diffusion (walk) begins. We also have the boundary conditions
$\tilde{c}=0$ at the two absorbing walls $x=0,1$, where the first
passage walks stop. Hence the solution is the normalized Green's
function

\begin{equation}
\tilde{c}(x,s) = \frac{\sinh \left({\sqrt{(s/D)}}x_<\right) \sinh
\left({\sqrt{(s/D)}}
(1-x_>)\right)}{\sqrt{(sD)}\sinh\left({\sqrt{(s/D)}}\right)}
\end{equation}

with $x_< = min(x,x_0)$ and $x_>= max(x,x_0)$.

The probability of passage to the walls (the eigenstates) is given
by the space derivatives with $s\rightarrow 0$.

\begin{eqnarray}
p(x=0) = D \left.\frac{ \partial \tilde{c}}{\partial
x}\right|_{s\rightarrow 0,x=0}= 1-x_0 = |a_1|^2   \nonumber\\
p(x=1) = -D \left. \frac{ \partial\tilde{c}}{\partial
x}\right|_{s\rightarrow 0,x=1} = x_0 = |a_0|^2
\end{eqnarray}

in conformity with quantum mechanics.

\section{WALKS IN HIGHER DIMENSIONAL HILBERT SPACES}

It is also possible to arrive at the above results using the two
variables $x$ and $y$ at all stages independently, and then
finally constraining them by the normalization condition $x+y=1$.
This provides a general procedure for an $n$-eigenvalue situation,
for arbitrary $n$. We shall illustrate the method for $n=3$, with
the initial (SD) state

\begin{equation}
|\psi\rangle_{SD} = |a_0|^2 |00^*\rangle_{SD} + |a_1|^2
|11^*\rangle_{SD} + |a_2|^2 |22^*\rangle_{SD}+ \ldots
\end{equation}

where we have labeled the three eigenvectors in an arbitrary
sequence, and as before we indicate the position of the initial
vector by $(x_0,y_0,z_0)$, and calculate the probabilities of
transitions to $x=1$, $y=1$ and $z=1$.

\begin{figure}[ht!]
\begin{center}
\includegraphics[width=10cm]{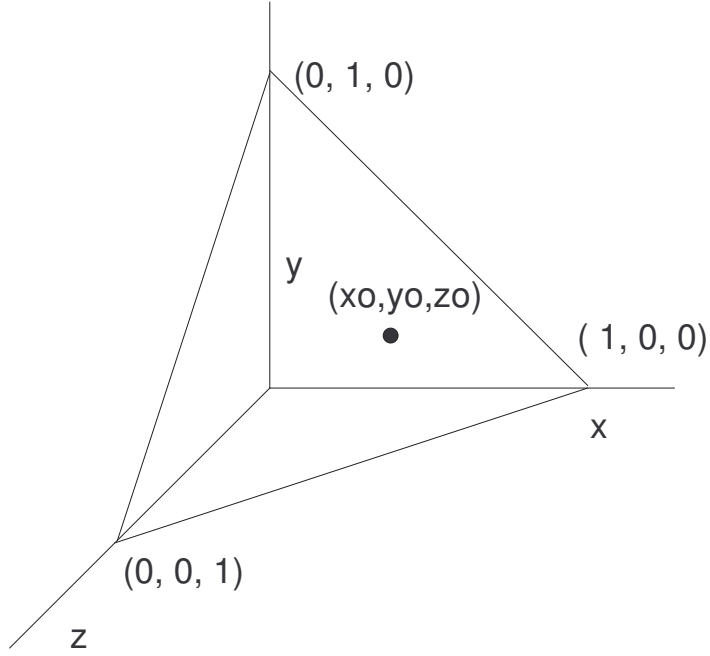}
\end{center}
\caption{\label{fig2}Three-dimensional Hilbert space with (SD)
eigenstate coefficients along the three axes. Walks begin at
$(x_0,y_0,z_0)$ and proceed on the triangle to any of the vertices
representing a pure eigenstate
 }
\end{figure}

 Fig.~\ref{fig2} shows the triangle in
which the co-ordinates are constrained, but for a symmetric
calculation in all three co-ordinates we shall impose the
normalization relation at the end. The diffusion equation is now

\begin{equation}
\nabla^2\tilde{c}({\bf x},s)- (s/D) \tilde{c}({\bf x};s)= -c({\bf
x},t=0)/D
\end{equation}

with the boundary condition

\begin{equation}
\left.\frac{\partial c}{\partial x_i}\right|_{x_i=0}=0
\end{equation}

which indicates zero diffusion out of the sides of the triangle of
Fig. 2. In game theory terms reaching $x_i=0$ eliminates $i$ from
the rest of the game.

Instead of using a Dirac delta function, we can normalize more
simply and with full symmetry in the co-ordinates by demanding
that at the source

\begin{equation}
\sum_i  (-D/2)\left( \left.\frac {\partial \tilde{c}}{\partial
x_i}\right|_{x_i=x_{0i}+\epsilon,s\rightarrow 0} + \left.\frac
{\partial \tilde{c}}{\partial
x_i}\right|_{x_i=x_{0i}-\epsilon,s\rightarrow 0} \right)= 1
\end{equation}

With symmetry among the co-ordinates, and hence the same velocity
of the walk (denoted by the parameter $k$ below) we finally get

\begin{equation}
\tilde{c}({\bf x}, s)= A \prod_i [\cosh(k x_{i<})] \cosh[k(2-
\sum_i x_{i>})]
\end{equation}

with $A$ given by normalization

\begin{equation}
A = \prod_i [\cosh (k x_{i0})] \sum_i \frac{\sinh[k (1-
x_{i0})]}{\cosh (k x_{i0})}
\end{equation}

and

\begin{equation}
k=\sqrt {\left(\frac{s}{3D}\right)}
\end{equation}

 This gives

\begin{eqnarray}
p_i = -D \left.\frac{\partial \tilde{c}}{\partial
x_i}\right|_{s\rightarrow 0, x_i=1}= x_i = |a_i|^2
\end{eqnarray}

which is a postulate in quantum theory.

For higher dimensions the procedure seems to be easily extensible,
with higher dimensional complexes in Hilbert space providing the
arena for the first passage diffusion. In each case we would need
the same boundary constraint for zero diffusion when a co-ordinate
goes to zero. The walk then proceeds in a lower dimensional
complex, till a vertex is reached.

\section{INTERPRETATIONAL IMPLICATIONS}
\subsection{Measuring Devices}
A measuring device records a particular state of an observable.  Hence, by quantum mechanical axiom, a measuring device is a macroscopic system that can indicate any one of the superposed quantum states and eliminate the rest, although the detailed mechanism of the process is not defined.
The manifestation of a highly  complex quantum state in the classical world of the device is a macroscopically  averaged truncation, that requires reorganization of a large number of smaller segments in favor of one effective quantum state to be expressed. The elimination of the remaining superposed components from being recorded in any measuring device  is denoted by a ``collapse" or a cancelation. This implies the impossibility of the remainder of the states of being coupled to another detector subsequently, in a manner as to be recordable. Because of the coupling of the measuring device with the environment, the recorded system's  surviving information is expressed within the S-D system.  The eliminated states are not expressed within S-D, but their characteristics disperse into the inactive components of the device that do not couple to the recorder, or into the external environment coupled to the device in a random manner.

\subsection{Coarse-graining and Macrostates}
Decoherence has been related to coarse graining in an attempt to explain probabilistic histories \cite{GE1}.  We extend that idea to claim that macroscopic expression of a state comes from its expression in a manner coherent enough to be expressible as reasonably sharp fuzzy sum (of states degenerate with respect to the characteristic measured, but with other attributes not necessarily agreeing unless constrained) at a certain scale. In order for a state to be expressible on a classical scale on a macroscopic recorder, a certain cluster of subsystems need to be expressed in a correlated manner within the system.  The transformation of expressed subclusters  into a correlated cluster expressing a single macrostate allows the entire cluster to influence other coupled clusters of the same scale. Just as the discrete  microscopic quantum spin ``up" and ``down" states exist, the macroscopically organized ``block spin" states may be thought to come in discrete assemblies as well that are in reality fuzzy sets of well-defined separated limits. For example, in biology, a hemoglobin can exist in one of two possible states depending on where or not an oxygen molecule is bound to it \cite{HAR}.

However, in the quantum domain, states can coexist as superpositions. The expression of a quantum component within the quantum scale (a small system) includes objects with clean wave functions with a minimal number of variables. The wavelike property of a small object can be observed when small (quantum scale) objects are allowed to pass through slits, and hence interfere. The interference reflects several possible states  existing in a superposed manner within the same microsystem's state function. In large ensembles, such wave-like properties are not usually coherently extended far enough to form quantum interactions with neighbors.

The macroscopic world is a large ensemble of interacting microsystems that does not display such wave properties. The presence of a large number of interacting overlapping neighbors effectively causes the decoherence of the individual microsystems' wave functions so that they become localized particle-like objects, quantum mechanically as well as in terms of classical mechanics. Such a set of interactions also eliminates bizarre combinations of dead and live cat wave functions, though valid paths exist for live cat states and dead cat states to migrate from one state to the other, quantum mechanically and classically, given sufficient time and energy.  Identity of the states of the macroscopic world are recognized not by quantum numbers or states of microscopic states, but rather by the correlation (and hence organization) of the subsystems.  A dead cat may also dissipate into dust after some time, signifying that a possible ensemble state in the scale of a macroscopic cat may dissipate into smaller ensembles and disperse into the uncorrelated large degrees of freedom within the environment in a manner such that the initial dead cat state cannot be retrieved from within the environment without an improbable conspiracy. Hence, the cat (a quantum mechanically fuzzy set) transforms from the live cat state to a dead cat state from its association with a detector which tangos with a superposed microsystem to either a dead cat associated with a fully decayed radioactive  or a still live cat associated with an intact harmless nucleus. We anticipate that the state of the detector is half-way between a fully quantized microsystem's and a completely decohered classical macroscopic recorder system, so that it has states pairing coherently with the microsystem on the one hand, and can also couple classically with other mesoscale systems in the recorder.

Hence, in our model, the entire universe does not get split into multiple energetically separate bands and multiple universes. The  critical interactions involved in the ``collapse" are carried out at the interface of the quantum and classical domains, introducing an intermediate meso-state.  The expression of a quantum state in the classical world is by means of a one-to-one correspondence between a quantum state and a macroscopic state, which might be a certain organization of the entire detector-detector ensemble.  As the entire detector is aligned in a certain direction, corresponding to  possible quantum states, information about the reduced components is lost to the rest of the recorder or the environment.

\subsection{Unitarity and Loss of Information}
In quantum computing, the operations are carried out by unitary operators, which are linear.  These operations preserve all information, so that it is possible to return to an initial state by means of an inverse function. Hence, the unitary operations map each quantum state into another state on a one-to-one basis.
An important theorem in quantum computing states that \cite{NC1}, it is not possible to clone an unknown state into a given state by means of unitary operations.  The proof is elegant and simple.  If two arbitrary initial states can be mapped to a single cloned states by means of a unitary operator, the inverse of that operator must yield both the initial states from the final cloned state, which violates the linearity clause.

However, in the case of quantum measurement, one of the possible superposed initial eigenstates is expressed in the detector, and the  probability of choice of the state is dependent on the amplitude. Hence, the quantum measurement process preserves partial information of the  state, and also can feel the wave amplitude.

The partial loss of information from the detector makes it clear that the stepwise linear unitary operators connecting the detector and the microsystem alone cannot cause measurement. The following alternate approaches might be able to explain quantum wave functions, so that the classical appearance of the world requires the two stages mentioned above:
\\
 {\bf a. Approximation of Waves}
 The perceived macroscopically identified universe may be
 taken to be the result of an extremely large number of random phase superpositions. The components of the detector contains systems  with a semilocal truncation near the detector that causes loss of information.

  The addition of a new superposed wave function in the large wave system causes a new approximation, which again makes some of the information in the original microscopic and isolated quantum wave function to be lost.  This approximation process may be summarized by taking a series of complicated unitary operators, and truncating the series, so that an effective non-linear operator arises, which is representative of the classical operation.
\\
  {\bf b. Energy Landscape}
  We propose an alternative explanation using interaction energy landscapes.
    In this scenario, an observer does not introduce a unitary operator with a known operation to an arbitrary state and expect it to evolve to a known (cloned) state.  Rather, quantum (many particle) mesoscale pairs exist in the detector, so that introducing a new incident quantum microsystem within the interaction range of another mesoscale system causes the two to interact and get coupled.  This process creates (virtually) bound pairs.  This explanation allows for cross terms to exist during the entire decoherence-reduction process.  The  active process involves interactions between two corresponding/similar states forming the matched active pairs involving  the micro and mesoscale levels, and the elimination of one state from the scenario automatically diminishes all the cross-terms passively.

    This picture does not need the existence of macroscopic environment states that are orthogonal and correspond to definite quantum states.  The initial (prior to reduction) existence of the entire set of orthogonal macro-states and the disappearance of all but one, would pose the problem of creating/annihilating large system states within the reduction time scale.  Also, the mechanism by which a microscopic quantum wave function is able to cause such a phenomenal situation is hard to explain.

Rather, in the energy landscape picture, pairs are formed at the lowest hierarchy level on a relatively small scale within the detector: sized between meso-subsystems and quantum microsystems. Internal couplings and interactions would then cause some microsystems within the detector to make a random walk to one quantum eigenstate together with the incident system, while the alternative small states are either rotated to favorable configurations, gaining the adaptation energy from the environment, or are dissipated within the environment by random phase cancelation, as there is no coherence glue to add up their contributions to the measured value, so that the level of expression of these ``lost states" are not coherently strong enough to indicate the ``non-collapsed" quantum states at a macroscopic level. This conversion can then trigger other similar coupled meso-systems within the detector-recorder complex, effectively amplifying the coupled walk at the lowest level.

In this approach, the entire ``universe" is broken down into quantum systems, mesosystems, detector components, detector and the rest of the environment, which is large, with many degrees of freedom, allowing some loss of information at each level of the scale hierarchy in the form of dissipation or leakage to external coupled components.  The initial interactions are local, but the dissipation of information allows for certain states to be coherently expressed within the highly coupled detector, while the attributes of the other states are dissipated within the large degrees of freedom of the environment.

\subsection{Choice of Basis}
While the outcome of the state in the detector is dependent on the quantum superposed waves interacting with the detector, the choice of basis is determined by the detector and the environment preceding it. For example, in the Stern-Gerlach experiment, the z-axis is determined when the electrons enter the apparatus.  The collapse of up states and down states take place along the z-axis.

This determination of the axis may be seen as innately built within the detector-environment complex.  A certain detector, by design, may have mesostates or subcomponents that bind with the quantum states only along that axis.  Hence, the detector may be seen as a system that is able to generate energetically favorable subsystems that can couple along only a certain axis.  There might be an ensemble of such states available, coupled with one another, which initially make the detector neutral.  But the introduction of the quantum state function causes the degenerate states to split and express themselves.

\section{CONCLUSIONS}

We believe the picture presented above is a simple, effective one
for understanding the process of collapse of superposed states to
eigenstates. One can also calculate the average time required for
the completion of the process trivially from the first passage
equations, and  it would depend on the parameter $D$ used as the
diffusion coefficient, which would vary according to the process
steps, as it would involve the details of the coupling between the
measuring device and the measured microsystem.

The question of nonlocality for the measurement  of entangled states
of spatially separated systems is an interesting one. In this
picture, and most others, the spatial separation has to be ignored
when entangled systems are considered, and the states can then be
processed in Hilbert space together with the corresponding measuring
devices. In \cite{FS2} we have shown recently that spatial
separation and local measurement of entangled systems are not
inconsistent with the established rules of quantum mechanics.

The randomness used here is not an inherent property of nature cited
in conventional quantum theory, but is simply the \emph{apparently}
unpredictable outcome of each of the steps representing the
interaction between an extremely large number of quantum components
in the device itself, and also of possible hidden parameters carried
from the source \cite{FS2}.

\section*{Acknowledgement}
The author would like to thank Prof Geoffrey Sewell for feedback, and for pointing out some possible changes. She would also like to thank Prof Ronald Omnes for reading an earlier draft.

\end{document}